\documentclass[conference]{IEEEtran}

\IEEEoverridecommandlockouts
\usepackage{listings}
\usepackage{makeidx}
\usepackage{algorithm,algpseudocode}
\usepackage{graphicx}
\usepackage{float}
\usepackage{cite}
\usepackage{amsmath}
\usepackage{rotating}
\usepackage{xspace}
\usepackage{multirow}
\usepackage{hyperref}
\usepackage{xspace}
\usepackage{paralist}
\usepackage{booktabs}
\usepackage{balance}
\usepackage{cleveref}
\usepackage{makecell}
\usepackage{color, colortbl}

\newcommand{\pp}[1]{\vspace{3pt}\noindent\textbf{\emph{#1}}}
\newcommand{\bp}[2]{\begin{tabular}{@{\textbullet~}p{#1}@{}}#2\end{tabular}}

\newcommand\blfootnote[1]{%
  \begingroup
  \renewcommand\thefootnote{}\footnote{#1}%
  \addtocounter{footnote}{-1}%
  \endgroup
}

\begin{document}

% Title
\title{A Community Roadmap for Scientific Workflows Research and Development}

% Authors
\author{
  \IEEEauthorblockN{
    Rafael Ferreira da Silva\IEEEauthorrefmark{1}\IEEEauthorrefmark{2},
    Henri Casanova\IEEEauthorrefmark{3},
    Kyle Chard\IEEEauthorrefmark{4}\IEEEauthorrefmark{5},
    Ilkay Altintas\IEEEauthorrefmark{6},
    Rosa M Badia\IEEEauthorrefmark{7},
    Bartosz Balis\IEEEauthorrefmark{8},\\
    Tain\~a Coleman\IEEEauthorrefmark{2},
    Frederik Coppens\IEEEauthorrefmark{9}\IEEEauthorrefmark{10},
    Frank Di Natale\IEEEauthorrefmark{11},
    Bjoern Enders\IEEEauthorrefmark{21},
    Thomas Fahringer\IEEEauthorrefmark{12},
    Rosa Filgueira\IEEEauthorrefmark{13},\\
    Grigori Fursin\IEEEauthorrefmark{14},
    Daniel Garijo\IEEEauthorrefmark{15},
    Carole Goble\IEEEauthorrefmark{16},
    Dorran Howell\IEEEauthorrefmark{17},
    Shantenu Jha\IEEEauthorrefmark{18},
    Daniel S. Katz\IEEEauthorrefmark{19},\\
    Daniel Laney\IEEEauthorrefmark{11},
    Ulf Leser\IEEEauthorrefmark{20},
    Maciej Malawski\IEEEauthorrefmark{8},
    Kshitij Mehta\IEEEauthorrefmark{1},
    Lo\"{i}c Pottier\IEEEauthorrefmark{2},
    Jonathan Ozik\IEEEauthorrefmark{4}\IEEEauthorrefmark{5},\\
    J. Luc Peterson\IEEEauthorrefmark{11},
    Lavanya Ramakrishnan\IEEEauthorrefmark{21},
    Stian Soiland-Reyes\IEEEauthorrefmark{16}\IEEEauthorrefmark{22},
    Douglas Thain\IEEEauthorrefmark{23},
    Matthew Wolf\IEEEauthorrefmark{1}
    \\
    % {\color{red}[Please Add your Name and Affiliation] (they will be sorted alphabetically by last name)}
  }
  \IEEEauthorblockA{
    \IEEEauthorrefmark{1}Oak Ridge National Laboratory, Oak Ridge, TN, USA \ \
    \IEEEauthorrefmark{2}University of Southern California, Marina Del Rey, CA, USA \\
    \IEEEauthorrefmark{3}University of Hawaii, Honolulu, HI, USA \ \
    \IEEEauthorrefmark{4}Argonne National Laboratory, Lemont, IL, USA \\
    \IEEEauthorrefmark{5}The University of Chicago, Chicago, IL, USA \ \
    \IEEEauthorrefmark{6}University of California, San Diego, La Jolla, CA, USA \\
    \IEEEauthorrefmark{7}Barcelona Supercomputing Center, Spain \ \
    \IEEEauthorrefmark{8}AGH University of Science and Technology, Krakow, Poland \\
    \IEEEauthorrefmark{9}Ghent University, Ghent, Belgium\ \
    \IEEEauthorrefmark{10}VIB Center for Plant Systems Biology, Belgium \\
    \IEEEauthorrefmark{11}Lawrence Livermore National Lab, Livermore, CA, USA \ \
    \IEEEauthorrefmark{12}University of Innsbruck, Innsbruck, Austria \\
    \IEEEauthorrefmark{13}Heriot-Watt University, Edinburgh, UK \ \
    \IEEEauthorrefmark{14}OctoML, USA \ \
    \IEEEauthorrefmark{15}Universidad Polit\'{e}cnica de Madrid, Spain \\
    \IEEEauthorrefmark{16}The University of Manchester, Manchester, UK \ \
    \IEEEauthorrefmark{17}Tweag, Z\"{u}rich, Switzerland \\
    \IEEEauthorrefmark{18} Brookhaven National Laboratory, Upton, NY, 11973 \ \
    \IEEEauthorrefmark{19}University of Illinois at Urbana-Champaign, USA \\
    \IEEEauthorrefmark{20}Humboldt-Universität zu Berlin, Berlin, Germany \ \
    \IEEEauthorrefmark{21}Lawrence Berkeley National Lab, Berkeley, CA, USA \\
    \IEEEauthorrefmark{22}University of Amsterdam, Amsterdam, The Netherlands\ \
    \IEEEauthorrefmark{23}University of Notre Dame, Indiana, USA \\
  }
}

\maketitle
\thispagestyle{empty}
\pagestyle{empty}

% Abstract
\begin{abstract}
The landscape of workflow systems for scientific applications is notoriously convoluted with hundreds of seemingly equivalent workflow systems, many isolated research claims, and a steep learning curve. To address some of these challenges and lay the groundwork for transforming workflows research and development, the WorkflowsRI and ExaWorks projects partnered to bring the international workflows community together. This paper reports on discussions and findings from two virtual ``Workflows Community Summits" (January and April, 2021). The overarching goals of these workshops were to develop a view of the state of the art, identify crucial research challenges in the workflows community, articulate a vision for potential community efforts, and discuss technical approaches for realizing this vision. To this end, participants identified six broad themes: FAIR computational workflows; AI workflows; exascale challenges; APIs, interoperability, reuse, and standards; training and education; and building a workflows community. We summarize discussions and recommendations for each of these themes.
\end{abstract}

% Keywords

% sections
\section{Introduction}

Scientific workflow systems are used almost universally across scientific domains for solving complex and large-scale computing and data analysis problems, and have underpinned some of the most significant discoveries of the past decades~\cite{badia2017workflows}. Many of these workflows have significant computational, storage, and communication demands, and thus must execute on a wide range of large-scale platforms, from local clusters over science or public clouds to upcoming exascale HPC platforms~\cite{ferreiradasilva-fgcs-2017}. Managing these executions is often a significant undertaking, requiring a sophisticated and versatile software infrastructure. 

Historically, many of these infrastructures for workflow execution consisted of complex, integrated systems, developed in-house by workflow practitioners with strong dependencies on a range of legacy technologies---even including sets of ad-hoc scripts. Due to the increasing need to support workflows, dedicated workflow systems were developed to provide abstractions for creating, executing, and adapting workflows conveniently and efficiently while ensuring portability.  While these efforts are all worthwhile individually, there are now hundreds of independent workflow systems~\cite{workflow-systems}. These are created and used by thousands of researchers and developers, leading to a rapidly growing corpus of workflows research publications. The resulting workflow system technology landscape is fragmented, which may present significant barriers for future workflow users due to the tens of seemingly comparable, yet usually mutually incompatible, systems that exist. 

\begin{table*}[!th]
\centering
\scriptsize
\caption{Summary of current workflows research and development challenges and proposed community activities.}
\vspace{-8pt}
\begin{tabular}{p{1.3cm}p{7.7cm}p{7.7cm}}
\toprule 
Theme & Challenges & Community Activities \\
\midrule
\makecell[l]{FAIR\\Computational\\Workflows} &
\bp{7.7cm}{
    Define FAIR principles for computational workflows that consider the complex lifecycle from specification to execution and data products
    \\
    Define metrics to measure the FAIRness of a workflow.
    \\
    Engage the community to define principles, policies, and best practices
}
&
\bp{7.7cm}{
    Review prior and current efforts for FAIR data and software with respect to workflows, and outline rules for FAIR workflows
    \\
    Define recommendations for FAIR workflow developers and systems
    \\
    Automate FAIRness in workflows by recording necessary provenance data
}
\\
\midrule

\makecell[l]{AI\\Workflows} &
\bp{7.7cm}{
    Lack of support for heterogeneity of compute resources and fine-grained data management features, versioning, and data provenance capabilities
    \\
    Lack of capabilities for enabling workflow steering and dynamic workflows
    \\
    Integration of ML frameworks into the current HPC landscape
}
&
\bp{7.7cm}{
    Develop comprehensive use cases for sample problems with representative workflow structures and data types
    \\
    Define a process for characterizing the challenges for enabling AI workflows
    \\
    Develop AI workflows as a way to benchmark HPC systems
}
\\
\midrule

\makecell[l]{Exascale\\Challenges\\and Beyond} &
\bp{7.7cm}{
    Resource allocation policies and schedulers are not designed for workflow-aware abstractions, thus users tend to use an ill-fitted job abstraction
    \\
    Unfavorable design of resource descriptions and mechanisms for workflow users/systems, and lack of fault-tolerance and fault-recovery solutions
}
&
\bp{7.7cm}{
    Develop documentation in the form of workflow templates/recipes/miniapps for execution on high-end HPC systems
    \\
    Specify benchmark workflows for exascale execution
    \\
    Include workflow requirements as part of the machine procurement process
}
\\
\midrule

\makecell[l]{APIs, Reuse,\\Interoperability,\\and Standards} &
\bp{7.7cm}{
    Workflow systems differ by design, thus interoperability at some layers is likely to be more impactful than others
    \\
    Workflow standards are typically developed by a subset of the community
    \\
    Quantifying the value of common representations of workflows is not trivial
}
&
\bp{7.7cm}{
    Identify differences and commonalities between different systems
    \\
    Identify and characterize domain-specific efforts, identify workflow patterns, and develop case-studies of business process workflows and serverless workflow systems
}
\\
\midrule

\makecell[l]{Training\\and Education} &
\bp{7.7cm}{
    Many workflow systems have high barrier to entry and lack training material
    \\
    Homegrown workflow solutions and constraints can prevent users from reproducing their functionality on workflow tools developed by others
    \\
    Unawareness of the workflow technological and conceptual landscape
}
&
\bp{7.7cm}{
    Identify basic sample workflow patterns, develop a community workflow knowledge-base, and look at current research on technology adoption
    \\
    Include workflow terminology and concepts in university curricula and software carpentry efforts
}
\\
\midrule

\makecell[l]{Building\\a Workflows\\Community} &
\bp{7.7cm}{
    Define what is meant by a ``workflows community''
    \\
    Remedy the inability to link developers and users to bridge translational gaps
    \\
    Pathways for participation in a network of researchers, developers, and users
}
& 
\bp{7.7cm}{
    Establish a common knowledge-base for workflow technology
    \\
    Establish a \emph{Workflow Guild}: an organization focused on interaction and good relationships and self-support between workflow developers and their systems
}
\\
\bottomrule
\end{tabular}
\label{tab:challenges}
\vspace{-10pt}
\end{table*}

In the current workflow research, there are conflicting theoretical bases and abstractions for what constitutes a workflow system. It may be possible to translate between systems that use the same underlying abstractions; however, the contrary is not feasible. Specifically, typical systems have a layered model that abstractly underlies it: (i)~if the models are the same for two systems, they are compatible to some extent, and if they implement the same layers, they can be interchanged (modulo some translation effort); (ii)~if the models are the same for two systems, but they are implemented by components at different layers, they can be complementary, and may have common elements that could be shared; (iii)~if the models are distinct, workflows or system components are likely not exchangeable or interoperable. As a result, many teams still elect to build their own custom solutions rather than adopt, adapt, or build upon, existing workflow systems. This current state of the workflow systems landscape negatively impacts workflow users, developers, and researchers~\cite{deelman2018future}.

The WorkflowsRI~\cite{workflowsri} and ExaWorks~\cite{al2021exaworks} projects have partnered to bring the workflows community (researchers, developers, science and engineering users, and cyberinfrastructure experts) together to collaboratively elucidate the R\&D efforts necessary to remedy the above situation. They conducted a series of virtual events entitled ``Workflows Community Summits'', in which the overarching goal was to (i)~develop a view of the state of the art, (ii)~identify key research challenges, (iii)~articulate a vision for potential activities, and (iv)~explore technical approaches for realizing (part of) this vision. The summits gathered over 70 participants from a group of international lead researchers and developers, from distinct workflow systems and user communities. The outcomes of the summits have been compiled and published in two technical reports~\cite{ferreiradasilva2021wcs, wcs2021technical}. In this paper, we summarize the discussions and findings by presenting a consolidated view of the state of the art, challenges, and  potential efforts, which we eventually synthesize into a community roadmap. Table~\ref{tab:challenges} presents, in the form of top-level themes, a summary of those challenges and targeted community activities. Table~\ref{tab:roadmap} summarizes a proposed community roadmap with technical approaches. 

The remainder of this paper is organized as follows. Sections~\ref{sec:fair}-\ref{sec:community} provide a brief state of the art and challenges for each theme and proposed community activities. Section~\ref{sec:roadmap} discusses technical approaches for a community roadmap. Section~\ref{sec:conclusion} concludes with a summary of  discussions.

\begin{table*}[!th]
\centering
\scriptsize
\caption{Summary of technical roadmap milestones per research and development thrust.}
\vspace{-8pt}
\begin{tabular}{p{2.8cm}p{14.4cm}}
\toprule 
Thrust & Roadmap Milestones \\
\midrule
\makecell[l]{Definition of common\\workflow patterns and\\benchmarks} &
\bp{14.4cm}{
    Define small sets of workflow pattern and benchmark deliverables, and implement them using a selected set of workflow systems
    \\
    Investigate automatic generation of patterns and configurable benchmarks (to enable weak and strong scaling experiments)
    \\
    Establish or leverage a centralized repository to host and curate patterns and benchmarks
}
\\
\midrule

\makecell[l]{Identifying paths toward\\interoperability of workflow\\systems} &
\bp{14.4cm}{
    Define interoperability for different roles, develop a horizontal interoperability (i.e., making interoperable components), and establish a requirements document per abstraction layer
    \\
    Develop real-world workflow benchmarks, use cases for interoperability, and common APIs that represent workflow library components
    \\
    Establish a workflow systems developer community
}
\\
\midrule

\makecell[l]{Improving workflow systems'\\interface with legacy and\\ emerging HPC software and\\hardware stacks} &
\bp{14.4cm}{
    Document a machine-readable description of key properties of widely used sites, and remote authentication needs from the workflow perspective
    \\
    Identify new workflow patterns (e.g. motivated from AI workflows), attain portability across heterogeneous hardware, and develop a registry of execution environment information
    \\
    Organize a community event involving workflow system developers, end users, authentication technology providers, and facility operators
}
\\
\bottomrule
\end{tabular}
\label{tab:roadmap}
\vspace{-10pt}
\end{table*}

\section{FAIR Computational Workflows}
\label{sec:fair}

The FAIR principles~\cite{wilkinson2016fair} have laid a foundation for sharing and publishing digital assets and, in particular, data. The FAIR principles emphasize machine accessibility and that all digital assets should be Findable, Accessible, Interoperable, and Reusable. Workflows encode the methods by which the scientific process is conducted and via which data are created. It is thus important that workflows both support the creation of FAIR data and themselves adhere to the FAIR principles.

%% Subsection
\subsection{Brief State-of-the-art and Challenges}

Workflows are hybrid processual digital assets that can be considered as data or software, or some combination of both. As such, there is a range of considerations to take into account with respect to the FAIR principles~\cite{goble2020}. Some perspectives are already well explored in data/software FAIRness, such as descriptive metadata, software metrics, and versioning; however, workflows create unique challenges such as representing a \textbf{\emph{complex lifecycle}} from specification to execution via a workflow system, through to the data created at the completion of the workflow execution.

As a specialized kind of software, workflows have two properties that FAIRness fundamentally must address: \textbf{\emph{abstraction and composition}}. As far as possible a workflow specification, as a graph or some declarative expression, is abstracted from its execution undertaken by a dedicated software platform. Workflows are composed of modular building blocks and expected to be remixed. FAIR applies ``all the way down" at the specification and execution level, and for the whole workflow and each of its components. One of the most challenging aspects of making workflows FAIR is ensuring that they can be \textbf{\emph{reused}}. These challenges include being able to capture and then move workflow components, dependencies, and application environments in such a way as not to  affect the resulting execution of the workflow. Further work is required to understand use cases for reuse, before exploring methods for capturing necessary context and enabling reuse in the same or different environments. 

Once use cases are defined, there are many \textbf{\emph{metrics and features}} that could be considered to determine whether a workflow is FAIR. These features may differ depending on the type of workflow and its application domain. Prior work in data and software FAIRness~\cite{wilkinson2016fair, katz2021taking} provides a starting point, however, these metrics need to be revised for workflows. In terms of labeling, there has been widespread adoption of reproducibility badges for publications and of FAIR labels for data in repositories~\cite{acm-badges}. Similar approaches could be applied to computational workflows. 
Finally, developing methods for FAIR workflows requires \textbf{\emph{community engagement}} (i)~to define principles, policies, and best practices to share workflows; (ii)~to standardize metadata representation and collection processes; (iii)~to create developer-friendly guidelines and workflow-friendly tools; and (iv)~to develop shared infrastructure for enabling development, execution, and sharing of FAIR workflows.

%% Subsection
\subsection{A Vision for Potential Community Activities}

Given current efforts for developing FAIR data and software, it is important to first understand what efforts could be adapted to workflow problems. An immediate activity include participating in working groups focused on applying FAIR principles to data and software. For instance, FAIR4RS~\cite{fair4rs} coordinates community-led discussions around FAIR principles for research software. Workflows could then be initially tackled from the point of view of workflows as software, which could originate a novel \textbf{\emph{task force}}. Proposed working groups such as FAIR for Virtual Research Environments~\cite{fair4vre} represent adequate progress towards this goal. 

A fundamental tenet of FAIR is the universal availability of machine processable metadata. The European EOSC-Life Workflow Collaboratory, for example, has developed a metadata framework for FAIR workflows based on schema.org~\cite{bioschemas-ComputationalWorkflow}, RO-Crate~\cite{rocrate}, and CWL~\cite{cwl}. This could be a community starting point for standardization of metadata about workflows.  

An integral aspect of a FAIR computational workflows task force would be to collect a set of real-world use cases and workflows in several domains to examine from the perspective of the FAIR data principles. This exercise will likely highlight areas in which the FAIR data principles adequately represent challenges in workflows. Based on these experiences, a set of \textbf{\emph{simple rules}} could be defined for creating FAIR workflows, similar to the ones in~\cite{monteil2020nine}. From these rules, 
prominent workflow repositories (e.g., WorkflowHub.eu~\cite{workflowhub} and Dockstore~\cite{yuen2021dockstore}), communities, and workflow systems can define \textbf{\emph{recommendations}} to support the development and sharing of FAIR workflows. These efforts relate not only to the workflows themselves, but the workflow components, execution environments, and the different types of data.

Ensuring \textbf{\emph{provenance}} can capture the necessary information is key for enabling FAIRness in workflows. Many provenance models~\cite{oliveira2018provenance} can be implemented or extended to capture the information needed for FAIR workflows. Additionally, FAIR principles are more likely to be followed if the process for capturing these metrics is automated and embedded in workflow systems. In this case, a workflow execution will become FAIR by default, or perhaps with minimal user curation.

\section{AI Workflows}
\label{sec:ai}

Artificial intelligence (AI) and machine learning (ML) techniques are becoming more and more popular within the scientific community. Workflows increasingly integrate ML models to guide analysis, couple simulation and data analysis codes, and exploit specialized computing hardware (e.g., GPUs, neuromorphic chips)~\cite{zhou2017machine}. These workflows inherently couple various types of tasks such as short ML inference, multi-node simulations, long-running ML model training, etc. They are also often iterative and dynamic, with learning systems deciding in real time how to modify the workflow, e.g. by adding new simulations or changing the workflow all together. AI-enabled workflow systems therefore must be capable of optimally placing and managing single- and multi-node tasks, exploit heterogeneous architectures (CPUs, GPUs, and accelerators), and seamlessly coordinate the dynamic coupling of disparate simulation tools.

%% Subsection
\subsection{Brief State-of-the-art and Challenges}

Workflows empowered with ML techniques largely differ from traditional workflows running on HPC machines. While workflows (i.e., one large-scale application simulating a scientific object or process) traditionally take little input data and produce large outputs, ML approaches target model training, which usually requires the input of a large quantity of data (either via files or from a collection of databases) and produces a small number of trained models. After training, these models are used to infer new quantities (during ``model inference") and behave like very lightweight applications that produce small quantities of output data. These models can be stand-alone applications, or even embedded within larger traditional simulations. There exists an inherent tension between traditional HPC, which evolved around executing large capacity-style codes and AI-HPC, which requires the coordinated execution of many smaller capability-scale applications (e.g., large ensembles of data generation co-mingled with inference and coinciding with periodic retraining of models).

With its reliance of data, effective AI-workflows should provide \textbf{\emph{fine-grained data management}} and versioning features, as well as adequate data provenance capabilities. This data management will have to be flexible: some applications and workflows might need to move data via a file-system, while others could be better served from a traditional database, data store, or a streaming dataflow model. During inference, it may be best to couple the (lightweight) model as close to the data it is processing as possible. In any case, effective data management is a key feature of successful AI workflows.

Another key feature of AI workflows is the inherent incorporation of non-traditional hardware, such as GPUs and tensor processing units (TPUs), which can significantly accelerate both training and inference steps. Workflow systems thus need to provide mechanisms for managing \textbf{\emph{heterogeneous resources}} -- offloading heavy computations to GPUs, and managing data between GPU and CPU memory hierarchies. Furthermore, since ML training and inference may be best executed on different hardware from the main simulation, AI workflows might need to be executed on \textbf{multi-machine} federated systems (with the main code executed on a traditional HPC system, but the ML model training or inference on a separate system). Additionally, it is also necessary to provide tight \textbf{\emph{integration to widely-used ML frameworks}}, the development of which is not driven by the HPC community. ML frameworks use Python and R-based libraries and do not follow the classic HPC model: C/C++/Fortran, MPI, and OpenMP, and submission to an HPC batch scheduler. Yet, some efforts in the HPC community seem promising, e.g. LBANN~\cite{lbann}, EMEWS~\cite{ozik2016, ozik_population_2021}, and eFlows4HPC~\cite{eflows}. Other approaches like Merlin~\cite{merlin} blend HPC and cloud technologies to enable federated workflows.  However, there is a clear disconnect between HPC motivations, needs, and requirements, and AI/ML current practices.

Finally, one of the major differences between traditional and AI workflows is the inherent \textbf{\emph{iterative nature of ML processes}} -- AI workflows often feature feedback loops over a data set. Data are created, the model is retrained, and its accuracy evaluated. ML training tasks might leverage hyperparameter optimization frameworks~\cite{akiba2019optuna, bergstra2013hyperopt, wozniak2018} to adjust their execution settings in real time. The final trained model is often used to select new data to acquire (in an ``active learning" environment, the model is used to decide which new simulations to run to better train the model on the next iteration). By design, ML-empowered workflows are dynamic, in contrast to traditional workflows with more structured and deterministic computations. At runtime, the workflow execution graph can potentially evolve based on internal metrics (accuracy), which may reshape the graph or trigger task preemption. Workflow systems should thus support \textbf{\emph{dynamic branching}} (e.g., conditionals, criteria) and partial workflow re-execution on-demand.

%% Subsection
\subsection{A Vision for Potential Community Activities}

To address the disconnect between HPC systems and practices and AI workflows, the community needs to develop sets of example \textbf{\emph{use cases for sample problems}} with representative workflow structures and data types. In addition to expanding upon the above challenges, the community could ``codify'' these challenges in example use cases. However, the set of challenges for enabling AI workflows is extensive. The community thus needs to define a \textbf{\emph{systematic process}} for identifying and categorizing these challenges. A short-term recommendation would be to write a ``community white paper" about AI Workflow challenges/needs.

Building from the use cases above for the needs and requirements of AI workflows, the community could define \textbf{\emph{AI-Workflow mini-apps}}, which could be used to pair with vendors/future HPC developers so that the systems can be benchmarked against these workflows, and therefore support the co-design of emerging or future systems (e.g., MLCommons~\cite{mlcommons} and the Collective Knowledge framework~\cite{fursin2020collective}).

\section{Exascale Challenges and Beyond}
\label{sec:exascale}

Given the computational demands of many workflows, it is crucial that their execution be not only feasible but also effortless and efficient on large-scale HPC systems, and in particular upcoming exascale systems~\cite{ferreiradasilva-fgcs-2017}. Exascale systems are likely to contain millions of independent computing elements that can be concurrently scheduled across more than 10,000 nodes, millions of cores, and tens of thousands of accelerators.

%% Subsection
\subsection{Brief State-of-the-art and Challenges}

HPC resource allocation policies and schedulers designs typically do not consider workflow applications: they provide a mere ``job" abstraction instead of workflow-aware abstractions. Workflow users/systems are forced to make their workflows run on top of this \textbf{\emph{ill-fitted abstraction}} -- e.g., it is difficult to control low-level behavior critical to workflows (i.e., precise mapping of tasks to specific compute resources on a compute node). Furthermore, there is a clear lack of support for elasticity (i.e., scaling up/down the number of nodes). Overall, it is currently difficult to run workflows efficiently and conveniently on HPC systems without extending (or even overhauling) resource management/scheduling approaches, which ideally would allow programmable, fine-grain application-level resource allocation and scheduling.

Related to the above challenge, it is currently not possible to support both workflow and non-workflow users harmoniously and/or efficiently on the same system. Some features needed by workflows are often unavailable. For instance, batch schedulers can support elastic jobs (e.g., Slurm); however, experience shows that system admins may not be keen on enabling this capability, as they deem long static allocations preferable. A \textbf{\emph{cultural change}} is perhaps needed as it seems that workflows are not yet considered as high-priority applications by high-end compute facilities.

Hybrid architectures are key to high performance and many workflows can or are specifically designed to exploit them. However, on HPC systems, the necessary \textbf{\emph{resource descriptions and mechanisms}} are not necessarily available to workflow users/systems (even though some workflow systems have successfully interfaced to such mechanisms on particular systems)~\cite{ahn2020flux}. Although these resource descriptions and mechanisms are typically available as part of the ``job" abstraction, it is often not clear how a workflow system can discover and use them effectively.

Finally, \textbf{\emph{fault-tolerance and fault-recovery}} have been extensively studied on exascale systems, with several works and working solutions for traditional parallel jobs~\cite{heldens2020landscape}. In the context of scientific workflows, specific techniques have been the subject of several studies~\cite{prathiba2017survey}, however workflow-specific solutions are typically not readily available or deployed. Moreover, workflows are built on smaller platforms, thus operating and testing at exascale would entail expressing new requirements/capabilities and dealing with new constraints (e.g., what is a ``local" exascale workflow?).

%% Subsection
\subsection{A Vision for Potential Community Activities}

An immediate activity consists in developing documentation in the form of \textbf{\emph{workflow templates/recipes/miniapps}} for execution on high-end HPC systems to be hosted on a community web site. Some efforts underway provide partial solutions~\cite{ewels2020nf}. For instance, collections of workflows exist but typically do not provide large scale execution capabilities (e.g., community testbeds). Some compute facilities provide workflow tool documentation and help with their users~\cite{nersc-workflows}. These solutions should be cataloged as a starting point, and HPC facilities could promote yearly ``workflow days", in which they give workflow users and developers training and early access to machines to try out their workflows, thus gathering feedback from users and developers.

To drive the design of workflow-aware abstractions, the community could specify \textbf{\emph{community benchmark workflows}} for exascale execution, exerting all relevant hardware as well as functionality capabilities. Then it becomes possible for different workflow systems to execute these benchmarks -- initial efforts could build on previous benchmark solutions~\cite{openebench, coleman2021wfcommons}. These benchmarks could then be included in \textbf{\emph{exascale machines acceptability tests}}. Note that there will be a need to pick particular workflow systems to run these benchmarks, which will foster training and education of HPC personnel.

Last, including workflow requirements very early on in \textbf{\emph{machine procurement process}} for machines at computing facilities will significantly lower the barriers for enabling workflow execution and therefore porting workflow applications. This effort is therefore preconditioned on the availability of miniapps and/or benchmark specifications, as well as API/scheduler specifications, as outlined above.

\section{APIs, Interoperability, Reuse, and Standards}
\label{sec:interoperability}

There has been an explosion of workflow technologies in the last decade~\cite{workflow-systems}. Individual workflow systems often serve a particular user community, a specific underlying compute infrastructure, a dedicated software engineering vision, or follow a specific historical trait. As a result, there are substantial technical and conceptual overlaps. Reasons for divergence include (i)~use cases require different workflow structures, (ii)~organizations have very different optimization goals, (iii)~predefined execution systems provide fundamentally different capabilities, or (iv)~availability and scarcity of different types of resources. Another reason is that it is relatively easy to start building a workflow system for a specific narrow focus (i.e., these systems have a gentle software development curve~\cite{SDCblog}), leading to large numbers of packages that provide some basic functionality, and developers who are subject to the sunk cost fallacy and then continue to invest in their custom packages, rather than joining forces and building community packages. This divergence leads to missed opportunities for interoperability. It is often difficult for workflows to be ported across systems, for system components to be interchanged, for provenance to be captured and exploited in similar ways, and for developers to leverage different execution engines, schedulers, or monitoring services.

%% Subsection
\subsection{Brief State-of-the-art and Challenges}

Workflow systems often grow organically: developers start by solving a concrete data analysis problem and they end up with a new workflow tool. In some cases, workflow systems may \textbf{\emph{differ by design}}, rather than by accident. For example, they offer fundamentally different abstractions or models for a workflow: DAG-structured \emph{vs.} recursive, imperative \emph{vs.} declarative, data flow \emph{vs.} control (and data) flow. These fundamental differences, catering for different use cases, make it such that full interoperability may simply not be possible. Alternately, workflow systems have many different layers and components that may be interchangeable, e.g., workflow specifications, task descriptions, data passing methods, file handling, task execution engines, etc. Interoperability at some layers is likely to be more impactful than others; for instance, being able to run the same workflow specification (with appropriately encapsulated task implementations) on different workflow infrastructures would be a major relief for users trying to reuse implemented workflows in other organizations. Further, interoperability does not need to imply agreement and for workflow systems to implement a standard interface; instead, it may occur via shim layers or intermediate representations, in a similar manner to compiling to a high level language. With the reuse goal, projects as eFlows4HPC proposes the HPC Workflows as a Service (HPCWaaS) methodology, where workflows will be defined by expert developers and provided as a service to community users~\cite{eflows}.

Most efforts to unify workflow systems and/or their components have led to the definition of a ``standard'' developed by a subset of the community~\cite{terstyanszky2014enabling, cwl-annotations}. However, the specialization of some of these standards may require that other systems conform to that specification, thus resulting in low adoption. Attempts to standardize also may lead to overly generic interfaces that in the end inhibit usability and lead to hidden incompatibilities. 

A particularly pressing problem at the interface of workflow technology and HPC systems is the need for a \textbf{\emph{common submission model}} that is compatible to heterogeneous platforms. The differences between the ways workflow engines, schedulers, and execution engines interact is a universal challenge faced by workflow developers when trying to target multiple infrastructures underlying long-lasting design decisions. Further challenges relate to authentication and authorization models deployed on many systems (e.g., two-factor authentication). Some efforts in this area are currently undergoing~\cite{JLESC-common-registry}.

%% Subsection
\subsection{A Vision for Potential Community Activities}

An immediate and continuous action would be to host several \textbf{\emph{``bake-offs''}} to compare workflow systems, including task and workflow definitions, a benchmark set of workflows with defined input data and outputs, as well as job execution interfaces. This would entail engaging participants to write and execute these workflows and identifying commonalities between systems. A successful example is the GA4GH-DREAM challenge~\cite{ga4gh}. An open question is whether such attempts should be domain-specific or domain-overarching;  there is likely a greater opportunity for standardization within domains (and indeed some domains have already made significant progress), but domain-specific standards would only partly solve the interoperability problem. The workflow community should then review these areas, determine and then publicize what has worked, and build on successful prior efforts.

With the emergence of FaaS (Function-as-a-Service) systems (e.g., AWS Lambda and Step Functions, Azure Durable Functions, Google Cloud Functions, IBM Composer), or CaaS (Container-as-a-Service) services (e.g. AWS Fargate, Google Cloud Run), the community should identify a set of \textbf{\emph{suggested use cases}} and compare them against an implementation with popular or recently developed FaaS-enabled workflow systems~\cite{chard2020funcx, smirnov2020apollo, malawski2020serverless}. Such a comparison may turn out complementary features that can be of benefit for both industry and the workflows community. In addition to features, a set of common workflow patterns could also be identified. However, there is still some uncertainty regarding the scope of previously developed patterns (e.g., for representing patterns in dynamic workflows). Thus, it is necessary to \textbf{\emph{survey published patterns}}~\cite{workflowpatterns,garijo2014common} and identify gaps seen by the community.

Although the above proposed activities have the potential to advance interoperability, the current funding and research recognition models often implicitly work against standardization by constantly requiring innovative ideas even in areas where outreach, uptake, and maintenance rather than innovation seems to be the most pressing problem. Developing \textbf{\emph{sustained funding models}} for building and evolving workflow standards, encouraging their adoption, supporting interoperability, testing, and providing user and developer training would help address these challenges. 

\section{Training and Education for Workflow Users}
\label{sec:training}

There is a strong need for more, better, and new training and education opportunities for workflow users. Many users ``re-invent the wheel" without reusing software infrastructures and workflow tools that would make their workflow execution more convenient, more efficient, easier to evolve, and more portable. This is partly due to the lack of comprehensive and intuitive training materials that would guide users through the process of designing a workflow (besides the typical ``toy" examples provided in tutorials).

%% Subsection
\subsection{Brief State-of-the-art and Challenges}

Using workflow tools can require large amounts of effort and time, due to a \textbf{\emph{steep learning curve}}. A contributing factor is that users may not know the required terminology and concepts. As a result, some have noted that what would be needed in the current technology landscape is to ``ship a developer along with the workflow tool''.

One of the reasons for the above challenge is that there are \textbf{\emph{few ``recipes'' or ``cookbooks''}} for workflow systems. Furthermore, given that workflows and their execution platforms are complex and diverse, in addition to mere training material, there is a need for a training infrastructure that consists of workflows and accompanying data (small enough to be used for training purposes but large enough to be meaningful) as well as execution testbeds for running these workflows.

Given the multitude of workflow systems~\cite{workflow-systems}, and the lack of standards (Section~\ref{sec:interoperability}), users cannot easily pick the appropriate systems for their needs. More importantly, there is an understandable fear of being locked into a tool that at some point in the near future will no longer be supported. Although documentation can be a problem, \textbf{\emph{guidance}} is the more crucial issue. Many users have the basic skills to create and execute workflows on some system, but as requirements gradually increase many users evolve their simple approaches in ad-hoc ways, thus developing/maintaining a working but \textbf{\emph{imperfect homegrown system}}. There is thus a high risk of hitting technological or labor-intensiveness roadblocks, which could be remedied by using a workflow system. But, when ``graduating" to such a system, there will likely be constraints that prevent users from reproducing the functionality of their homegrown system. The benefits of using the workflow system should thus largely outweigh the drawback of these constraints.

Given all the above challenges, it is not easy to \textbf{\emph{reach out to users at the appropriate time}}. Reach out too early and users will not view using a particular workflow system as compelling. Reach out too late, and users are already locked into their homegrown system, even though in the long run this system will severely harm their productivity.

%% Subsection
\subsection{A Vision for Potential Community Activities}

Lowering the entry barrier is key for enabling the next-generation of researchers to benefit from workflow systems. An initial approach would be to provide a basic set of simple, yet conceptually rich, \textbf{\emph{sample workflow patterns}} (e.g., ``hello world" one-task workflows, chain workflows, fork-join workflows, simple dynamic workflows), all with a few ways of handling data and I/O, and all with a few target execution platforms. Then workflow system teams can provide (interactive) documentation (or could be hosted on a community Web site) on how to run these patterns with their system~\cite{nersc-workflows}. Additionally, mechanisms should be identified at the institutional level to commit workflow systems \textbf{\emph{training efforts in person}}: (i)~this should be based on existing facilities and universities efforts; (ii)~the scope of the training should be narrowed down so it is manageable; and (iii)~the issue of ``who trains the trainers?" needs to be addressed. 

In light of workforce training, workflow concepts should be taught at early stages of the researchers/users education path. Precisely, these concepts should be included in university curricula, including domain science curricula. Recent efforts have produced pedagogic modules that target workflow education~\cite{casanova2021eduwrench, eduwrench}. Pedagogic content could also be distributed as workflow modules to existing software carpentry efforts~\cite{swcarpentry}.

There is an established community of workflow researchers, developers, and users that has extensive expertise knowledge regarding specific tools, systems, applications, etc. It is crucial to capture such knowledge and bootstrap a \textbf{\emph{community workflow knowledge-base}} (following standards for documentation, interoperability, etc.) for training and education. The workflows community would also benefit from collaborations with social scientists and sociologists so as to help define an overall strategy for approaching some of the above challenges.

\section{Building a Workflows Community}
\label{sec:community}

Given the current large size and fragmentation of the workflow technology landscape, there is a clear need to establish a cohesive community of workflow developers and users. This community would be crucial for avoiding unnecessary duplication of effort and would allow for sharing, and thus growing, of knowledge. To this end, there are four main components that need to be addressed for building a community: (i)~identity building, (ii)~trust, (iii)~participation, and (iv)~rewards.

%% Subsection
\subsection{Brief State-of-the-art and Challenges}

The most natural idea is to think of two \textbf{\emph{distinct communities}}: (i)~a Workflow Research and Development Community, and (ii)~a Workflow User Community. The former gathers people who share interest in workflow R\&D, and corresponding sub-disciplines. Subgroups of this community are based on common methodologies, technical domains (e.g., computing, provenance, design), scientific disciplines, as well as geographical and funding areas. The latter gathers anyone using workflows for optimization of their work processes. However, most domain science users think of themselves in their specific disciplines first, as they just happen to use workflows to get their work done.

The two aforementioned communities are not necessarily disjoint, but currently have little overlap. And yet, it is crucial that they interact. Such interaction seems to happen only on a case-by-case basis, rather than via organized community efforts.  One could, instead, envision a single community (e.g., team of users, or ``team-flow'') that gathers both workflow system developers and workflow-focused users, with the common goal of spreading knowledge and adoption of workflows, thus working towards increased \textbf{\emph{sharing and convergence/interoperation}} of technologies and approaches.

Establishing trust and processes is key for bringing both communities together. There is no one-size-fits-all workflow system or solution for all domains, instead each domain presents their own specific needs and have different preferred ways to address problems. There is a pressing need for \textbf{\emph{maintaining documentation and dissemination}} that fits different usage options and needs.

%% Subsection
\subsection{A Vision for Potential Community Activities}

Given the above, there are a number of existing community efforts that could serve as inspiration, e.g., the WorkflowHub Club~\cite{workflowhub} and Galaxy~\cite{galaxy}. One approach is to gather experience from computing facilities where teams have successfully adopted and are successfully running workflow systems~\cite{nersc-workflows}. Another possibility is to use proposal/project reviews as mechanisms for spreading workflow technology knowledge. Specifically, finding ways to make proposal authors (typically domain scientists) aware of available technology would prevent their proposed work to not entail re-inventing the wheel. Finally, it is clear that solving the ``community challenge'' has large overlap with solving the ``education challenge'' (Section~\ref{sec:training}).

A short-term activity would entail \textbf{\emph{establishing a common knowledge-base for workflow technology}} so that workflow users would be able to navigate the current technology landscape. User criteria (for navigation) need to be defined. Workflow system developers can add to this knowledge base via self-reporting and could include test statuses for a set of standard workflow configurations, especially if workflow systems are deployed across sites. There is large overlap with similar proposed community efforts identified in Sections~\ref{sec:exascale} and~\ref{sec:training}.

An ambitious vision would be to \textbf{\emph{establish a ``Workflow Guild''}}, i.e., an organization focused on interaction and good relationships and self-support between subscribing workflow developers and their systems, as well as dissemination of best-practices and tools that are used in the development and use of these systems. However, there are still barriers to be conquered: (i)~such a community could be too self-reflecting, and yet still remain fragmented; (ii)~a cultural/social problem is that creating a new system is typically more exciting for computer scientists as opposed to re-using someone's system; and (iii)~building trust and reducing internal competition will be difficult, though building community identity will help the Guild work together against external competitors.

\section{A Roadmap for Workflows Research and Development}
\label{sec:roadmap}

In the previous sections, we have listed broad challenges for the workflows community and proposed a vision for community activities to address these challenges. Here, we explore technical approaches for realizing (part of) that vision. Based on the outcomes of the first summit~\cite{ferreiradasilva2021wcs}, we identified three technical thrusts for discussion in the second summit~\cite{wcs2021technical}. Some of these thrusts align with a single theme of the first summit and some are cross-cutting. In the following subsections, we present the summary of discussions at the second summit and propose roadmap milestones that emerged from these discussions. Additional details can be found in~\cite{wcs2021technical}, and a summary of the roadmap milestones is shown in Table~\ref{tab:roadmap}.

%% Subsection
\subsection{Defining Common Workflow Patterns and Benchmarks}

The above sections point to strong needs for establishing repositories of common workflow patterns and benchmarks (Sections~\ref{sec:ai}, \ref{sec:exascale}, \ref{sec:interoperability}, and~\ref{sec:training}). One objective is to develop workflow patterns in which each pattern should be easy for users to leverage as starting point for their own specific workflow applications -- they should provide links to one or more implementations, where each implementation is for a particular workflow system and can be downloaded and easily modified by the user. However, the \emph{level of abstraction} of these patterns (i.e., the level of connection to real application use-cases) should still be defined. At one extreme, workflow patterns could be completely abstract with no connection to any real-world application. At the other extreme, workflow patterns could be completely use-case-driven and correspond to actual scientific applications, with realistic task computations and data sets. The end goal is thus to identify useful patterns that span the spectrum of possible levels of abstraction. 

Another aspect is the level of detail with which a pattern specifies the platform on which it is to be executed and the logistics of the execution on that platform. The platform description could be left completely abstract, or it could be fully specified. Under-specifying execution platforms and logistics may render the pattern not useful, but over-specifying them could render the pattern too niche.

Benchmark specifications should make it easy for workflow system developers to develop them or to determine that their system cannot implement these specifications. Each benchmark should provide links to implementations and data sets, where each implementation is for a particular workflow system. These implementations would be provided, maintained, and evolved by workflow system developers. They should be able to be packaged so that they are executed out of the box on the classes of platforms they support. Moreover, the input data of these workflows should be configurable in size to enable both weak and strong scaling experiments. For all configurations, also the output of the workflow must be provided to allow functional testing.

Given the above, the following milestones are proposed:

\pp{M1.}
Define small sets (between 5 and 10) of workflow patterns and of workflow benchmark deliverables. These should be defined by eliciting feedback from users and workflow system developers, as well as based on existing sources that provide or define real-world or synthetic workflow patterns. 

\pp{M2.}
Work with a selected set of workflow systems to implement the above patterns and benchmarks.

\pp{M3.}
Investigate options for automatic generation of patterns and/or benchmarks using existing approaches~\cite{katz2016application, coleman2021wfcommons}.

\pp{M4.}
Identify or create a centralized repository to host and curate the above patterns and benchmarks~\cite{workflowhub, coleman2021wfcommons}.

%% Subsection
\subsection{Paths Toward Interoperability of Workflow Systems}

Workflow systems differ at varying degrees such as expressivity, execution models, and ecosystems. These differences are mainly due to individual implementations of language, control mechanisms (e.g., fault tolerance, loops), data management mechanisms, execution backends, reproducibility aspects for sharing workflows, and provenance and FAIR metadata capturing. The need for interoperability is paramount and can happen at multiple technical levels (e.g., task, tools, workflows, data, metadata, provenance, and packaging) as well as non-technical level including semantics, organizational, and legal issues (e.g., licenses compatibility, data sharing policies).

The need for interoperability of workflow applications and systems is commonly modeled as a problem of porting applications across systems, which may require days up to weeks of development effort~\cite{schiefer2020portability,LFB+21}. Most of the previous approaches for tackling the interoperability problem attempted to develop complete vertical solutions. However, there is no attempt to develop an approach from a perspective of making interoperable components. Interoperable components require standardized APIs, which are still an open challenge~\cite{turilli2019middleware, billings2017toward}. 

There is a tendency to tie the workflow with its execution model and data structures (e.g, the intertwine between the abstract workflow and its execution). Understanding which component in the workflow system architecture accounts for which functionality, is then paramount. Thus, separation of concerns is key for interoperability at many levels, e.g. separation of orchestration of the workflow graph from its execution.

Given the above, the following milestones are proposed:

\pp{M5.}
Define concrete notions of interoperability for different stakeholders, in particular workflow designers, workflow system designer, and workflow execution organizations.

\pp{M6.}
Establish a ``requirements" document per a small set of \emph{abstraction layers} that will (i)~capture the commonalities between components of workflow systems; and (ii)~perform a separation of concerns to identify interoperability gaps.

\pp{M7.}
Develop real-world workflow \emph{benchmarks} featuring different configurations and complexities (see previous section). Such benchmarks would be key to evaluate the functionality of workflow systems and computing platforms systematically.

\pp{M8.}
Develop \emph{use cases} for interoperability based on real-life scenarios, e.g., porting workflows across platforms that would provide different file system and/or different resource manager.

\pp{M9.}
Develop \emph{common APIs} that represent a set of workflow library components, so as interoperability could be achieved at the component level~\cite{cwl, arshad2015definition, Fursin_2021}, including APIs for defining inputs, storing intermediate results, and output data.

\pp{M10.}
Establish a workflow systems \emph{developer community}. An immediate activity would be to develop a centralized repository of workflow-related research papers, and a workflow system registry aimed at DevOps and/or users.

%% Subsection
\subsection{Improving Workflow Systems' Interface with Legacy and Emerging HPC Software and Hardware Stacks}

Improving the interface between workflow systems and existing as well as emerging HPC and cloud stacks is particularly important as workflows are designed to be used for long periods of time and may be moved between computing providers. This challenge is  exacerbated with the specialization of hardware and software systems (e.g., with accelerators, virtualization, containers, and cloud or serverless infrastructures). Thus, it is important to address the challenges faced by workflow systems with respect to discovering and interacting with a diverse set of cyberinfrastructure resources and also the difficulties authenticating remote connections while adhering to facility policies.

Workflow systems require a standard method for querying a site on how to use that site, for example, information about the batch system, file system configuration, data transfer methods, and machine capabilities. It is crucial then to first understand what information is needed by workflow systems, what information could be made available programmatically and what would need to be manually curated (similar ongoing efforts~\cite{sgci} may provide the foundations for this effort).

A key capability provided by workflow systems is remote job execution, which is necessary in cases where workflows span facilities. However, authentication has always been challenging. Many workflow systems rely on fragile SSH connections and in the past the use of GSISSH for delegated authentication. Recently, sites have moved towards two factor authentication and even OAuth-based solutions. There are though ongoing efforts to provide programmatic identity and access management in scientific domains~\cite{alt2020oauth, withers2018scitokens}. While the topic of remote authentication is much more broad than the workflows community, there are important considerations that should be included in this discussion related to programmatic access, community credentials, and long-term access. 

Given the above, the following milestones are proposed:

\pp{M11.}
Document a machine-readable description of the essential properties of popular sites, e.g., define a JSON schema and share it on GitHub.

\pp{M12.}
Document remote authentication requirements from the workflow perspective and organize an event involving workflow system developers, end users, authentication technology providers, and facility operators. 

\section{Conclusion}
\label{sec:conclusion}

In this paper, we have documented and summarized the wealth of information acquired as a result of a series of virtual events entitled the ``Workflows Community Summit". The goal of these summits was to identify the common and current challenges faced by the workflows community, and outline a vision for short- and long-term community activities. From this vision, we have defined a community roadmap consisting of 12 milestones, which proposes solutions and technical approaches for achieving that vision. This initial series of successful events bespoke the need for continued engagement among workflow researchers, developers, and users, as well as enlarging the scope of the community to also embrace key stakeholders (e.g., computing facility operators, funding agency representatives, etc.) for enabling the proposed vision and roadmap.

\blfootnote{
\noindent\textbf{Acknowledgments.}
This work was funded by NSF awards \#2016610, \#2016682, and \#2016619, and by the Exascale Computing Project (17-SC-20-SC), a collaborative effort of the US DOE and the NNSA.
This research used resources of the Oak Ridge Leadership Computing Facility at the Oak Ridge National Laboratory, which is supported by the Office of Science of the U.S. DOE under Contract \#DE-AC05-00OR22725.
CG and SSR acknowledge funding from European Commission's contracts BioExcel-2 (H2020-INFRAEDI-2018-1 823830), EOSC-Life (H2020-INFRAEOSC-2018-2 824087), IBISBA 1.0 (H2020-INFRAIA-2017-1-two-stage 730976), PREP-IBISBA (H2020-INFRADEV-2019-2 871118), SyntheSys+ (H2020-INFRAIA-2018-1 823827). UL acknowledges funding from the German Research Council for CRC 1404 FONDA.
FC is supported by the Research Foundation-Flanders (FWO, I002819N) and by the European Union's Horizon 2020 research and innovation programme under grant \#824087 (EOSC-Life).
BSC authors acknowledge EuroHPC JU under contract 955558 (eFlows4HPCproject).
We thank all participants of the Workflows Community Summits, held in January 2021 and April 2021.
}

% bibliography
\bibliographystyle{IEEEtran}

% \balance

\end{document}